\documentclass[useAMS,usegraphicx,usenatbib]{mn2e}
\usepackage{url}
\usepackage{amssymb}
\newcommand{\myr}{{q}}
\newcommand{\myin}{{i}}
\newcommand{\myout}{{o}}
\newcommand\apj{{ApJ}}% 
\newcommand\apjl{{ApJ}}% 
\newcommand\apjs{{ApJS}}% 
\newcommand\mnras{{MNRAS}}% 
\newcommand\physrep{{Phys.~Rep.}}% 
\begin{document}

\title[Shock Induced Cold Fronts]{Cold Fronts from Shock Collisions}

\author[Birnboim Keshet \& Hernquist]{Yuval Birnboim$^{1}$, Uri Keshet$^{1}$\thanks{Einstein fellow} \& Lars Hernquist$^{1}$\\
$^{1}$Harvard-Smithsonian Center for Astrophysics, 60 Garden Street,
  Cambridge MA, USA}
\date{Accepted ---. Received ---; in original ---}

\pagerange{\pageref{firstpage}--\pageref{lastpage}} \pubyear{2010}

\bibliographystyle{mn2e}
\maketitle

\label{firstpage}

\begin{abstract}

Cold fronts (CFs) are found in most galaxy clusters, as well as in
some galaxies and groups of galaxies.  We propose that some CFs are
relics of collisions between trailing shocks.  Such a collision
typically results in a spherical, factor $\approx 1.4-2.7$
density/temperature discontinuity.  These CFs may be found as far out
as the virial shock, unlike what is expected in other CF formation
models.

As a demonstration of this effect, we use one dimensional simulations
where halo reverberations involving periodic  
collisions between the virial shock and outgoing secondary shocks exist.  
These collisions yield a distinctive, concentric geometric sequence of
CFs which trace the expansion of the virial shock.

\end{abstract}

\begin{keywords}
galaxies: clusters: general  -- galaxies: haloes -- X-rays: galaxies: clusters -- shock waves
\end{keywords}

\section{Introduction}
%origin of CFs
Recent X-ray observations of galaxy clusters reveal various phenomena
in the gaseous haloes of clusters, such as mergers, cavities, shocks
and cold fronts (CFs).  CFs are thought to be contact discontinuities,
where the density and temperature jump, while the pressure
remains continuous (up to projection/resolution effects).  They are
common in clusters \citep{markevitch07} and vary in morphology (they
are often arcs but some radial or filamentary CFs exist), contrast
(the density jump is scattered around a value of $\sim 2$), quiescence
(some are in highly disturbed merging regions, and some in smooth,
quiet locations), and orientation.  CFs have been postulated to
originate from cold material stripped during mergers \citep[][ for
  example]{markevitch00} or sloshing of the intergalactic medium
\citep[IGM; ][]{markevitch01,ascasibar06}.
%Additional Properties of CFs
In some cases, a metallicity gradient is observed across the CF,
indicating that it results from stripped gas, or radial motions of gas
\citep[see ][and references therein]{markevitch07}.  Shear is often
found along CFs in relaxed clusters, implying nearly sonic bulk flow
beneath the CF \citep{keshet10}.  Whether
CFs are present at very large radii ($\gtrsim 500~{\rm kpc}$) is
currently unknown because of observational limitations.

In what follows, we propose that some CFs are produced by collisions
between shocks.  When two trailing shocks collide, a CF is always
expected to form.  Its parameters can be calculated by solving the
shock conditions and corresponding Riemann problem.  Most of the
scenarios in which shocks are produced (quasars, AGN jets, mergers)
predict that shocks will be created at, or near, a cluster center, and
expand outwards (albeit not necessarily isotropically).  When one
shock trails another, it always propagates faster (supersonically with
respect to the subsonic downstream flow of the leading shock), so
collisions between outgoing shocks in a cluster are inevitable if they
are generated within a sufficiently short time interval.

In \S~\ref{sec:riemann} we solve for the parameters of CFs that are
caused by a collision between two arbitrary planar shocks. As a
demonstration of the general mechanism, we show in \S~\ref{sec:virial}, using 1D
spherical cluster simulation that collisions between the virial shock
and a secondary shock are expected to produce a distinct pattern of
CFs. 
In \S~\ref{sec:observe} we compare the
features of observed CFs with results from our analysis, and list some
model-specific predictions.  In \S~\ref{sec:summary} we summarize and
conclude.

\section{Collision Between Two Trailing Shocks}
\label{sec:riemann}

In this section we examine two planar shocks moving in the same
direction and calculate the contact discontinuity that forms when the
second shock overtakes the leading one.  This problem is fully
characterized (up to normalization) by the Mach numbers of the first
and second shocks, $M_0$ and $M_1$.  At the instant of collision, a
discontinuity in velocity, density and pressure develops,
corresponding to a Riemann problem \citep[second case
in][\S93]{landau59}.  The discontinuity evolves into a (stronger)
shock propagating in the initial direction and a reflected rarefaction
wave, separated by a contact discontinuity which we shall refer to as
a shock induced CF (SICF).  The density is higher (and the entropy
lower) on the SICF side closer to the origin of the shocks.  This
yields a Rayleigh-Taylor stable configuration if the shocks are
expanding outwards.  Now, we derive the discontinuity parameters.

\subsection{The discontinuity contrast}

We consider an ideal gas with an adiabatic constant $\gamma.$ Before
the shocks collide, denote the unshocked region as zone $0$, the
region between the two shocks as zone $1$, and the doubly shocked
region as zone $2$.  After the collision, regions $0$ and $2$ remain
intact, but zone $1$ vanishes and is replaced by two regions, zone
$3_\myout$ (adjacent to zone $0$; the outer region for outgoing
spherical shocks) and zone $3_\myin$ (adjacent to $2$; inner),
separated by the SICF.  We refer to the plasma density, pressure,
velocity and speed of sound respectively as $\rho$, $p$, $u$ and $c$.
Velocities of shocks at the boundaries between zones are denoted by
$v_i$, with $i$ the upstream zone.  We rescale all parameters by the
unshocked parameters $\rho_0$ and $p_0$.

The velocity of the leading shock is
\begin{eqnarray}
v_0&=&u_0+M_0~c_0,\label{eq:v0}
\label{eq:v}
\end{eqnarray}
where
\begin{eqnarray}
c_i&=&\left(\frac{\gamma p_i}{\rho_i}\right)^{1/2}
\end{eqnarray}
for each zone $i$.  Without loss of generality, we measure velocities
with respect to zone $0$, implying $u_0=0$.

The state of zone $1$ is related to zone $0$ by the Rankine-Hugoniot conditions,
\begin{eqnarray}
p_1&=&p_0\frac{2\gamma M_0^2-\gamma+1}{\gamma+1}, \label{eq:purho1}\\
u_1&=& u_0 + \frac{p_1-p_0}{\rho_0(v_0-u_0)}, \label{eq:purho2}\\
\rho_1&=&\rho_0\frac{u_0-v_0}{u_1-v_0}. \label{eq:purho3}
\label{eq:purho}
\end{eqnarray}
The state of the doubly shocked region (zone $2$) is related to zone
$1$ by reapplying equations \ref{eq:v}-~\ref{eq:purho} with the
subscripts $0,1$ replaced respectively by $1,2$, and $M_0$ replaced by
$M_1$.

Across the contact
discontinuity that forms as the shocks collide (the SICF), the
pressure and velocity are continuous but 
the density, temperature and entropy are not; the CF contrast is
defined as $\myr\equiv\rho_{3\myin}/\rho_{3\myout}$.  Regions $0$ and
$3_\myout$ are related by the Rankine-Hugoniot jump conditions across
the newly formed shock,
\begin{eqnarray}
p_3&=&p_0\frac{(\gamma+1)\rho_{3\myout}-(\gamma-1)\rho_0}{(\gamma+1)\rho_0-(\gamma-1)\rho_{3\myout}}, \label{eq:u2u0a}\\
u_3-u_0&=&\left[(p_3-p_0)\left(\frac{1}{\rho_0}-\frac{1}{\rho_{3\myout}}\right)\right]^{1/2} \,. \label{eq:u2u0b}
\end{eqnarray}
The adiabatic rarefaction from pressure $p_2$ down to $p_3=p_{3\myin}=p_{3\myout}$ is determined by
\begin{eqnarray}
u_2-u_3&=&-\frac{2c_2}{\gamma-1}\left[1-\left(\frac{p_3}{p_2}\right)^{(\gamma-1)/2\gamma}\right] . \label{eq:u2u0c}
\end{eqnarray}
The system can be solved by noting that the sum of
eqs. (\ref{eq:u2u0b}) and (\ref{eq:u2u0c}) equals $u_2-u_0$, fixing
$\rho_{3\myout}$ as all other parameters are known from
eqs. (\ref{eq:v0}-\ref{eq:u2u0a}).

Finally, the Mach number of the new shock and the rarefacted density are given by
\begin{eqnarray}
M_f^2 & = & \frac{2\rho_{3\myout}/\rho_0}{(\gamma+1)-(\gamma-1)\rho_{3\myout}/\rho_0} \,, \\
\rho_{3\myin}&=&\rho_2(p_3/p_2)^{1/\gamma} .
\end{eqnarray}

\begin{figure}
\includegraphics[width=3.5in]{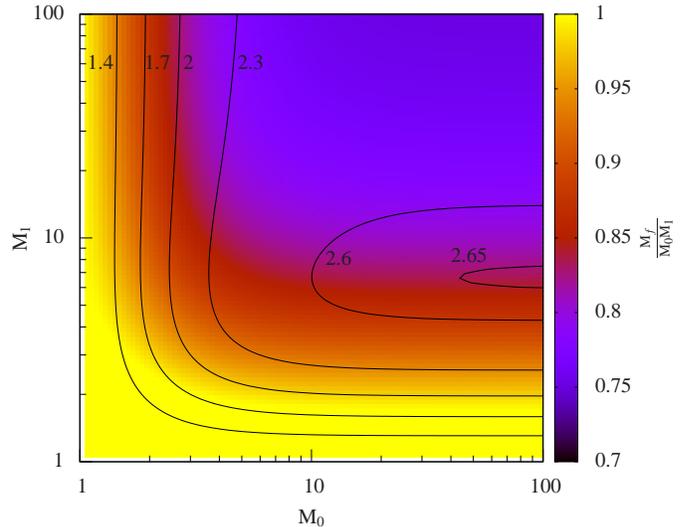}
\caption{Contrast $\myr$ of a CF generated by a collision between two trailing
  shocks with Mach numbers $M_0$ and $M_1,$ for $\gamma=5/3$ (labeled contours).
  The color map shows the final Mach number $M_f$ normalized by $M_0~M_1.$\label{fig:m0m1}}
\end{figure}
Figure \ref{fig:m0m1} shows the discontinuity contrast $q$ for various
values of $M_0$ and $M_1,$ for $\gamma=5/3.$  Typically $\myr\sim 2,$
ranging from $1.45$ for two Mach $2$ shocks, for example, to
$\myr_{max}$.  The maximal contrast $\myr_{max}$ depends only on the
adiabatic index; for $\gamma=5/3$ we find $\myr_{max}=2.653$, achieved
for $M_0\gg 1$ and $M_1\simeq 6.65$.  The possible contrast range of
SICFs is in good agreement with the CF contrasts observed
\citep{markevitch07,owers09}.  The figure colorscale shows the
dimensionless factor $f$ defined by $M_f=f M_0 M_1$, ranging between
$0.75$ and $1$ for $1<\{M_0,M_1\}<100$.

\subsection{Stability of Shock Induced Cold Fronts}
\label{sec:stability}

The stability of CFs limits the time duration over which they are
detectable, and so is important when comparing CF formation models
with observations.  Various processes can cause a gradual breakup or
smearing of the CF.  They act in CFs regardless of their formation
mechanism.

Thermal conduction and diffusion of particles across the CF smears the
discontinuity on a timescale that is set by the thermal velocity and
the mean free path of the protons. The Spitzer m.f.p. $\lambda$ in an
unmagnetized plasma with typical cluster densities is a few kpc, and
depends on the thermal conditions on both sides of the discontinuity
\citep{markevitch07}. Taking $\lambda \sim 10~{\rm kpc}$ and a thermal
velocity $\bar{v}\sim 1000~{\rm km \, sec}^{-1},$ and assuming that a
CF is visible if it is sharper than $L_{\rm obs}\sim 10~{\rm kpc},$ we
get a characteristic timescale for CF dissipation,
\begin{equation}
t\sim \frac{L_{\rm obs}^2}{D}\sim\frac{3L_{\rm
    obs}^2}{\bar{v}\lambda}\sim 10^7{\rm yr}.
\end{equation}
This result indicates that in order for shock induced CFs to be
observable, either (i) they are formed frequently \citep[e.g., by a
series of AGN bursts; see][]{ciotti07,ciotti09}; or (ii) magnetic
fields reduce the m.f.p. considerably, as some evidence suggests
\citep[see the discussion in][]{lazarian06}.

Note that the heat flux driven buoyancy instability (HBI) tends to
preferentially align magnetic fields perpendicular to the heat flow in
regions where the temperature decreases in the direction of gravity.
This effect could reduce radial diffusion within the inward cooling
regions in the cores of cool core clusters
\citep[][]{quataert08,parrish09}.  Across a CF transition there is a
sharp temperature drop towards the inner side, in the direction of
gravity.  Heat flow could induce HBI on the scale width of the CF,
aligning the magnetic fields parallel to the discontinuity and
diminishing radial diffusion.  This possibility needs to be addressed
further by detailed numerical magneto-hydrodynamic simulations.

\begin{figure}
\includegraphics[width=3.3in]{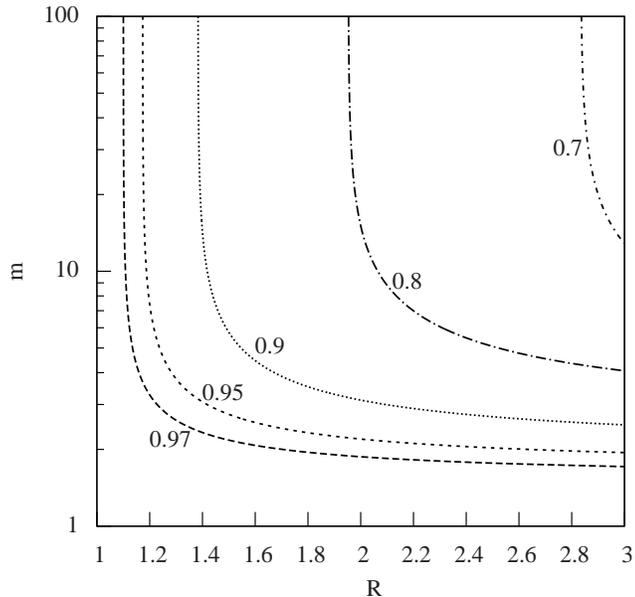}
\caption{Fractional decline (contours) in initial CF contrast $\myr$
  induced by a collision with a shock of Mach number $m$ arriving from the dense CF side, for $\gamma=5/3.$
  The vertical dashed line corresponds to the maximal SICF contrast, $\myr_{max}$.
  \label{fig:mr}}
\end{figure}
CFs could be degraded by subsequent shocks that sweep outwards across
the CF.  However, the CF contrast is only slightly diminished by this
effect, by $<4\%$ for $\myr < 2$ and a shock Mach number $m<2$, as
shown in fig.~\ref{fig:mr}.  Such passing shocks seed
Richtmyer-Meshkov instabilities which could cause the CF to break
down.  Although less efficient than Rayleigh-Taylor instabilities,
these instabilities operate regardless of the alignment with gravity.
The outcome depends on $m$ and on any initial small perturbations in
the CF surface.  However, the instability is suppressed if the CF
becomes sufficiently smoothed.

It is worth noting that KHI, which could break down CFs formed through
ram pressure stripping and sloshing, does not play an important role
in SICFs because there is little or no shear velocity across them.  On
the other hand, the stabilizing alignment of magnetic fields caused by
such shear \citep{markevitch07} is also not expected here.

\section{A Cluster Scale Reverberation Mode}
\label{sec:virial}

At the edge of galaxy clusters there are virial shocks with typical
Mach numbers $\sim 30-100,$ heating the gas from $\sim 10^4~{\rm K}$
to $\sim 10^7~{\rm K}$ by converting kinetic energy into thermal
energy.
The rate of expansion of a virial shock is set on average by the
mass flux and velocity of the infalling material
\citep[see for example][]{bertschinger85}.
We find in our simulations that while this expansion is
steady on average, the shock sometimes oscillates,
moving faster or slower than the steady state expansion.
This creates periods dominated by alternating halo compression and expansion.
During periods of enhanced compression in the outer halo regions, when the shock slows down, compression is sent inwards and is reflected through the center, gradually steepening into an outgoing shock.
When this shock collides with the virial shock, a situation corresponding to that described in \S~\ref{sec:riemann}, bouncing the virial shock into another cycle of reverberation, creating an SICF and sending inwards a steep rarefaction\footnote{A movie of the evolution of radial
profiles in time is available at
\url{http://www.cfa.harvard.edu/~ybirnboi/SICF/sicf.html} }.
This rarefaction marks the end of the compression period of the next
cycle setting the oscillation period to approximately the sound
crossing time, in and out across the cluster. The combined effect is a
long-lived, coherent ``breathing'' mode. 
Such ``breathing'' oscillations are observed in the 3D galactic and cluster halo simulations of \citet{kh09}\footnote{Du{\v s}an Kere{\v s},
(private communication).}.

Using 1D spherical simulations of the formation of clusters from
initial cosmic perturbations \citep{bd03}, we test the formation and
evolution of associated SICFs.  The simulation includes baryonic
shells, dark matter shells, and angular momentum support.  The
oscillatory mode in the simulation has not been intentionally excited,
and results from stochastic inner core ($\sim 0.05R_{vir}$) vibrations as
external dark matter shells with low angular momentum interact
gravitationally with the core.
Physically, any waves, shocks or gravitational perturbations
(mergers or AGNs) will contribute to the mode,
as well as non-smooth accretion rates that would cause the virial
shock to vibrate.  A more detailed examination of this is beyond the
scope of this letter (note, however, that the strength of the SICF
depends only weakly on the parameters of the secondary shock and is
always close to $2.$)

\begin{figure}
\includegraphics[width=3.3in]{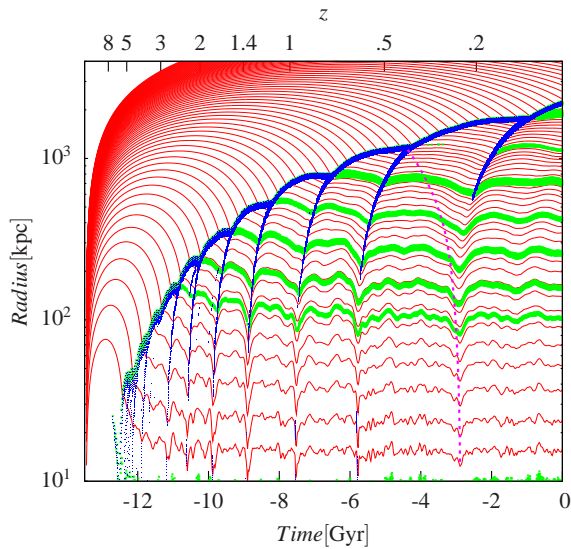}
\caption{Evolution of a galaxy cluster from cosmological initial
  perturbations. The final mass of the cluster at $z=0$ is $3\times
  10^{14}M_\odot.$ The radius and time of Lagrangian shells (every
  $25^{th}$ shell) are  plotted in red thin lines. Shocks are traced by large
  Lagrangian derivatives of the entropy ($d{\rm ln}S/dt>.1~{\rm Gyr}^{-1}$,
  blue dots), and CFs are traced by their large entropy gradients
  ($\partial{\rm ln} S/\partial {\rm ln}r>0.5$, green dots).
  Rarefaction waves can be seen as small motions of the Lagrangian
  shells (illustrated by the dashed magenta curve, manually added
  based on a time series analysis). Their trajectory is approximately
  the reflection of the preceding outgoing compression, time inverted
  about the last secondary-virial shock collision.
  \label{fig:time}}
\end{figure}

\begin{figure}
\includegraphics[width=3.6in]{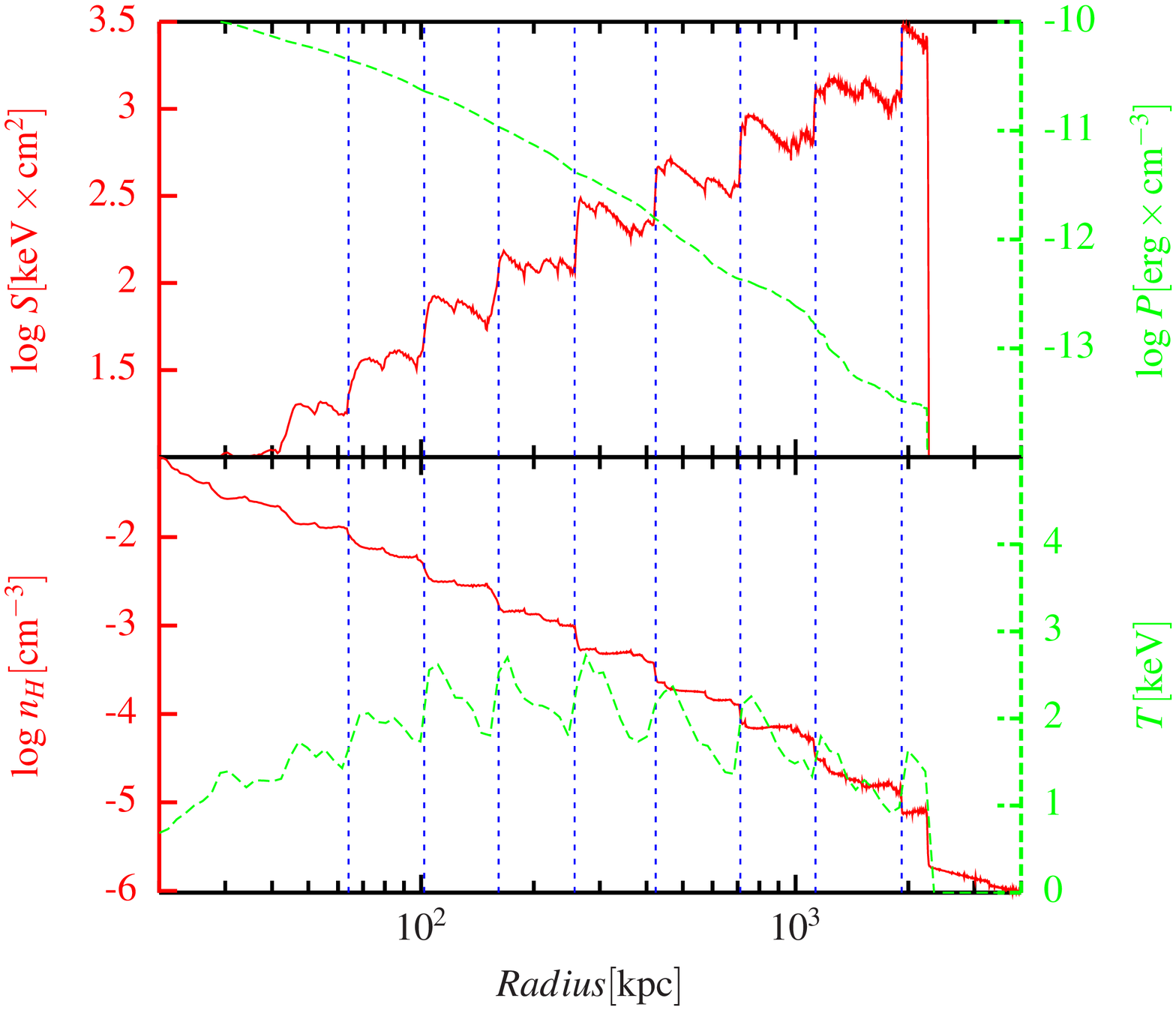}
\caption{Thermodynamic profile of the simulated cluster shown in Fig.~(\ref{fig:time}) at $z=0.$ {\it Top:}
  entropy (red, solid, left axis) and pressure (green, dashed, right
  axis). {\it Bottom:} density (red, solid, left axis) and temperature
  (green, dashed, right axis). CFs are marked with blue dashed
  vertical lines.\label{fig:prof}
  }
\end{figure}
Fig.~\ref{fig:time} shows the evolution of a $3\times
10^{14}M_{\odot}$ halo until redshift $z=0$ (an adiabatic version of
those described in Birnboim \& Dekel 2009, {\it in prep.}). The
expanding virial shock is traced by a jump in the velocity of the
Lagrangian shells (red, thin lines), and by the shock-finding
algorithm that is based on entropy increase along Lagrangian shells
(blue region).  The outgoing, secondary shocks that reach the virial
shock periodically are also clearly visible.  The virial shock bounces
outwards each time it is hit by a secondary.  A rarefaction wave is
reflected inwards (illustrated by a dashed curve), and a spherical
SICF is left behind. The SICFs are visible through the entropy
gradients (green region), and, in the absence of diffusion and
instabilities, persist until $z=0.$  As expected, the interaction of
the discontinuity with subsequent shocks (\S \ref{sec:stability};
fig.~\ref{fig:mr}) does not considerably reduce the CF contrast.
Fig.~\ref{fig:prof} shows the entropy, pressure, density and
temperature profiles of this simulation at $z=0.$  A series of SICFs
(marked with blue vertical lines) is visible, seen as entropy jumps
with continuous pressure.  The density jumps by roughly a factor of
$q\sim 2$, as expected, with the temperature dropping accordingly.

The SICFs formed in this simulation are almost static, and roughly
logarithmically spaced.  The calculations presented in Figures
\ref{fig:time}-\ref{fig:prof} are adiabatic.  When cooling is turned on
in the absence of feedback, this simulation suffers from overcooling,
creating an overmassive BCG of $2\times 10^{12}M_\odot$ and star
formation rates of $100~M_\odot{\rm yr}^{-1}$, and the luminosity
exceeds the $L_x-T$ relation. The excitation of the basic mode, and
periodic CFs, are observed in all these simulations. The simulation
presented here was performed with $2,000$ baryonic and $10,000$
dark matter shells. When the resolution is lowered to $250$ baryonic
shells, the reverberation amplitude and periods are essentially
unchanged, indicating that the results are well converged.

\section{Observability of Shock Induced Cold Fronts}
\label{sec:observe}
\citet{owers09} present high quality Chandra observations of
$3$ relaxed CFs. They all seem concentric with respect to the cluster
center, and spherical in appearance. \citet{owers09} interpret these as
evidence for sloshing, and find spiral characteristics in $2$ of
them.  We argue here that some CFs of this type could originate from
shock collisions.  In the SICF model, CFs are spherical, unlike the
truncated CFs formed by ram pressure stripping or sloshing, so no
special projection orientation is required.  On the other hand, SICFs
do not involve metallicity discontinuities, observed in some of these
CFs.  The observed contrasts of these CFs are quite uniform, with best
fit values in the range $2.0-2.1$ (with $\sim 20\%$ uncertainty) in
all three, as expected for SICFs\footnote{Note that collisions with
outgoing shocks could diminish the contrast towards $q\sim 2$,
regardless of CF origin.  A larger, more complete sample of CFs is
needed to determine the origin/s of the relaxed population.}.

The SICF formation model is entirely distinct from ram pressure
stripping and sloshing CFs, and it makes many different predictions.

{\bf Morphology.} SICFs are quasi-spherical around the source of the
shocks. A galactic merger defines an orbit plane, and the
corresponding perturbation (stripped material or center displacement)
will create CFs parallel to this plane.  In some sloshing
scenarios \citep{ascasibar06} the CFs extend considerably above the
plane, but would never appear as a full circle on the sky; an observed
CF ``ring'' would strongly point towards an SICF.  The statistical
properties of a large CF sample could thus be used to distinguish
between the different scenarios.

{\bf Amplitude.} SICFs have distinct entropy and density contrasts
that depend weakly on shock parameters; $q$ is typically larger than
$\sim 1.4$ (assuming $M_1\geq 2$) and is always smaller than
$q_{max}=2.65$.

{\bf Extent.} If the cluster reverberation mode is excited, SICF radii
should be approximately logarithmically spaced.
Any shock that expands and collides with the virial shock will create
SICF at the location of the virial shock, far beyond the core.
The external SICFs are younger, and would appear sharper owing to less
diffusion and collisions with secondary shocks.  Deep observations capable of
detecting CFs at $r\sim 1~{\rm Mpc}$ are predicted to find SICFs
(fig.~\ref{fig:prof}).  Such distant CFs occur naturally only in the
SICF model.

{\bf Plasma diagnostics.}  Shocks are known to modify plasma
properties in a non-linear manner, for example by accelerating
particles to high energies and amplifying/generating magnetic
fields. The plasmas on each side of an SICF may thus differ, being
processed either by two shocks or by one, stronger shock.  This may
allow indirect detection of the CF, in particular if the two shocks
were strong before the collision.  For example, enhanced magnetic
fields below the CF may be observable as excess synchrotron emission
from radio relics that extend across the CF, in nearby clusters, using
future high-resolution radio telescopes (LOFAR, SKA).

\section{Summary and Discussion}
\label{sec:summary}

Our study consists of two parts. The first is a general discussion
about CFs that form as a result of a collision between two trailing
shocks. The second part describes a specific reverberation mode of
galaxy and cluster haloes that has been identified in 1D simulations, and
also seen in 3D. One aspect of these reverberations is periodic
collisions between the virial shock and outgoing secondary shocks,
resulting in SICFs -- a special case of the mechanism discussed in the
first part of our paper.

We have shown that when shocks move in the same direction they collide
and generate a CF. The density contrast across the CF is calculated as
a function of the Mach numbers of the two shocks. It is typically
larger than $1.4$ (if $M\gtrsim2$), and is always smaller than
$q_{max}=2.65$. The discontinuity is smeared over time by diffusion,
at a rate that depends on the unknown nature and amplitude of magnetic
fields.
CFs are susceptible to heat-flux-driven buoyancy instability (HBI),
which could align the magnetic field tangent to the CF and potentially
moderate further diffusion. SICFs, like all other CFs, are subject to
Richtmyer-Meshkov instabilities from subsequent shocks passing through
the cluster. Such collisions reduce the CF contrast until it reaches
$q\sim 2$.  Unlike most other CF models, an SICF is not expected to
suffer from KHI.

In \S~\ref{sec:virial}, using a 1D spherical hydrodynamic code to
evolve cosmological perturbations, we demonstrate that a reverberation
mode exists in the haloes of galaxies and clusters  
and causes periodic collisions between the virial shocks and secondary
shocks that produce SICF. The simulated SICF contrast is
consistent with the theoretical predictions of \S~\ref{sec:riemann}.
A more thorough investigation of this potentilly important mode,
including its stability in 3D is left for future works. We use it here
as a specific scenario for SICF formation.

The predicted properties of SICFs are presented in
\S\ref{sec:observe}, and reproduce some of the CF features discussed
in \citet{owers09}.  In particular, we suggest that CFs in relaxed
clusters, with no evidence of mergers, shear, or chemical
discontinuities, may have formed by shock collisions.  We list the
properties of SICFs that could distinguish them from CFs formed by other
mechanisms.
The SICF model predicts quasi-spherical CFs which are concentric about
the cluster center, with contrast $\myr \sim 2,$
and possibly extending as far out as the virial shock.
An observed closed (circular/oval) CF could only be an SICF.
In the specific case of cluster reverberation,
a distinct spacing pattern between CFs is expected.
It may be possible to detect them indirectly, for example as
discontinuities superimposed on peripheral radio emission.

Shocks originating from the cluster center naturally occur in feedback
models that are invoked to solve the overcooling problem.  They are
also formed by mergers of substructures with the BCG.  Thus, SICFs
should be a natural phenomenon in clusters.  Further work is needed to
assess how common SICFs are with respect to other types of CFs, and to
characterize inner SICFs that could result, for example, from
collisions between offset AGN shocks.  The properties of reverberation
in 3D will be pursued in future work.

\section*{Acknowledgements}

We thank M. Markevitch for useful discussions.  YB acknowledges the
support of an ITC fellowship from the Harvard College Observatory.  UK
acknowledges support by NASA through Einstein Postdoctoral Fellowship
grant number PF8-90059 awarded by the Chandra X-ray Center, which is
operated by the Smithsonian Astrophysical Observatory for NASA under
contract NAS8-03060.

\label{lastpage}
\end{document}